# Observation of superconductivity in bilayer graphene/hexagonal boron nitride superlattices

Satoshi Moriyama[1,*], Yoshifumi Morita[2,*], Katsuyoshi Komatsu[1], Kosuke Endo[1,3], Takuya Iwasaki[4], Shu Nakaharai[1], Yutaka Noguchi[3], Yutaka Wakayama[1], Eiichiro Watanabe[5], Daiju Tsuya[5], Kenji Watanabe[6], Takashi Taniguchi[6].

**Affiliations:**

[1]International Center for Materials Nanoarchitectonics (WPI-MANA), National Institute for Materials Science (NIMS), Tsukuba, Ibaraki 305-0044, Japan.

[2]Faculty of Engineering, Gunma University, Kiryu, Gunma 376-8515, Japan.

[3]School of Science & Technology, Meiji University, Kawasaki 214-8571, Japan.

[4]International Center for Young Scientists (ICYS), NIMS, Tsukuba, Ibaraki 305-0044, Japan.

[5]Nanofabrication Platform, NIMS, Tsukuba, Ibaraki 305-0047, Japan.

[6]Research Center for Functional Materials, NIMS, Tsukuba, Ibaraki 305-0044, Japan.

*Corresponding to: MORIYAMA.Satoshi@nims.go.jp (S.M.); morita@gunma-u.ac.jp (Y.M.).

**Abstract:**

A class of low-dimensional superconductivity (SC), such as most "atomic-layer" SCs, has survived only under certain circumstances, implying a role of the substrate. Moreover, in some recent SC discoveries at heterogeneous interfaces, SC was buried in bulk solids and *ex situ*. Genuine atomic-layer SC is difficult to access. Here we report a novel route to atomic-layer SC in graphene superlattices. Our device comprises stacked non-twisted bilayer graphene (BLG) and hexagonal boron nitride (hBN), i.e., hBN/BLG/hBN Moiré superlattices. Upon *in situ* electrostatic doping, we observe an SC dome with a critical temperature up to $T_{BKT}$ = 14 K, corresponding to the confinement of vortices. We believe that SC via doping Dirac materials is ubiquitous in condensed matter and that this study paves a way toward the design of a new SC family.

**Main Text:**

Superconductivity (SC) has been one of the central topics in condensed matter physics since the discovery of SC phenomena and theory [*1, 2*]. In particular, the discovery of atomic-layer superconductors will have consequences for both fundamental physics and applications and implies a novel route to high critical temperature ($T_c$) SC, such as in cuprates [*3*] and quantum information devices. The emergence of Dirac fermions in solids ("Dirac materials") has been well established since the discovery of the "1$^{st}$ generation" in graphene [*4, 5*]. Very recently, SC due to doping a "Mott" insulator has been reported in magic-angle twisted bilayer graphene (BLG) superlattices [*6*]. In the early stage, fine tuning to a "magic" angle with vanishing velocity/flat band was focused upon. However, more relaxed conditions are later found to be sufficient for the energy dispersion/band width, and the role of other possible key ingredients, e.g., large density of states (DoS)/van Hove singularity (vHs), have been recently suggested (see, for example, ref.'s [*7, 8*]).

The field of SC in graphite intercalations has a long history [*9, 10*], and SC in carbon materials has long been sought after with a promise of high yields from both fundamental and application points of view. In this context, carbon-based superlattices are a novel class of quantum metamaterials. In particular, graphene superlattices comprise vertically stacked ultra-thin/atomic-layer quasi-two-dimensional materials, which distinctly differ from conventional molecular beam epitaxy/pulse laser deposition (MBE/PLD)-grown superlattices [*11*].

Herein, we report a novel route to atomic-layer SC in graphene superlattices via *in situ* electrostatic on/off switching. We note that fine tuning to a magic angle is not necessary in our device. Moreover, the ability to switch between a superconducting state and a "parent state" [*12, 13*] opens a door to state-of-art engineering in atomic-layer quantum devices. Further, small carrier concentration/Fermi pockets should lead to SC with enhanced critical fluctuations [*14–16*] and intriguing phenomena.

Our device is fabricated by stacking BLG and hexagonal boron nitride (hBN) (hBN/BLG/hBN stacks) with a small angle near zero between one of the two hBN sheets and the BLG (BLG itself is non-twisted); this is called a Moiré superlattice. This Moiré superlattice serves as a stage for our demonstration. Via *in situ* doping, we observe tunable zero resistance states.

In this study, we employed hBN/BLG/hBN Moiré superlattices. Fig. 1(A) shows a schematic of the typical structure of our device. In the optical microscope image, two devices are shown. In this paper, we employed the larger of the two. The resistance is defined through the four-terminal resistance $R_{ij,kl}$, which is defined by the voltage drop between terminals $k$ and $l$ divided by the electrical current injected between $i$ and $j$ in Fig. 1(A) (see also the Methods Summary section for more details). Fig. 1(B) shows an intensity map of the longitudinal resistance, $R_{xx}$, as a function of the back-gate voltage, $V_g$, and the magnetic field, $B$ (applied perpendicular to the substrate), at 6 K. In Fig. 1(B) also shows the correspondence between $V_g$ and density $n$. We estimated $n$ through our device structure/electrostatic capacitances (see also the Supplementary Information for the definition of $n$). In these superlattices, a long wavelength Moiré pattern occurs and leads to a Hofstadter butterfly under a magnetic field [17, 18]. Graphene Moiré superlattices have recently been intensively studied, in particular, Moiré bands/butterflies in BLG [18–20]. A Moiré superlattice leads to an energy gap at the charge neutral point (CNP), at which $n = 0$ cm$^{-2}$, and the emergence of satellites of the CNP. When subject to a magnetic field, the resistance peaks lead to 1st and 2nd generation Landau fans. The 1st generation corresponds to the CNP. The 2nd generation is due to inversion-symmetry breaking by hBN and corresponds to the satellites of the CNP. Fig. 1(B) also shows the Landau fans. The longitudinal and Hall resistivities exhibit basically the same pattern as seen in previous reports [18]; the pronounced peak in the longitudinal resistance at the CNP occurs at a gate voltage, $V_g \sim 0$ V, and the satellite resistance peak occurs at $V_g \sim -30$ V, which is referred to as "satellite" for simplicity. When subject to a magnetic field, these resistance peaks lead to the 1st and 2nd generation Landau fans, respectively (Fig. 1(B)). The alignment angle between the graphene and hBN is estimated to be $\theta \sim 0°$ [13]. Further, the measurement shows a Landau level formation with Hall conductance ($\sigma_{xy}$) steps of $4e^2/h$, where some degeneracies are lifted and additional plateaus also occur (Fig. 1(C)). These Quantum Hall effect results are the characteristics of BLG. Fig. 1(D) shows $R_{xx}$ as a function of $n$ at various temperatures without a magnetic field ($B = 0$ T). Sudden drop in $R_{xx}$ is observed around $n \sim -3.5 \times 10^{12}$ cm$^{-2}$, which indicates a precursor to zero resistances of SC. The inset of Fig. 1(D) provides $R_{xx}$ as a function of temperature, $T$, at the satellite ($n = -1.93 \times 10^{12}$ cm$^{-2}$), and the optimal doping for SC ($n = -3.48 \times 10^{12}$ cm$^{-2}$). At the optimal doping, SC shows the highest transition temperatures as shown in Fig. 2. At the satellite, the resistance shows non-metallic behavior due to the hBN-induced band gap [18]. Fig. 1(E) shows typical resistances, $R_{xx}$, as a function of temperature, $T$, without a magnetic field. At the lowest temperature, data show low resistance below the noise floor, corresponding to the regime with sudden resistance drop in Fig. 1(D). The I–V characteristics are shown in Fig. 1(F) for various temperatures near optimal doping ($n = -3.59 \times 10^{12}$ cm$^{-2}$), which shows SC critical current behavior at low temperatures (see the Supplementary Information for details of the transition temperatures analysis).

Fig. 2 shows the resistance, $R_{xx}$, as a function of both density, $n$, and temperature, $T$, without a magnetic field. A dome-shaped superconducting region, an SC dome, appears in the phase diagram. Inside the dome, data show SC behavior as shown in Fig. 1(E, F). The SC appears sharply at $n \sim -3.2 \times 10^{12}$ cm$^{-2}$ and $-3.6 \times 10^{12}$ cm$^{-2}$. The SC critical temperature saturates near optimal doping ($n \sim -3.48 \times 10^{12}$ cm$^{-2}$), which leads to the dome-shaped SC phase referred to as an SC dome. We observe no (correlated) insulator behavior between the satellite and SC dome. Upon *in situ* electrostatic doping, we observe an SC dome with a critical temperature $T_{BKT} = 14$ K (see the Supplementary Information for details of the transition temperatures analysis).

Fig. 3(A) shows the magnetoresistance, $R_{xx}(B)$, with focus on the regime near $n \sim -3.5 \times 10^{12}$ cm$^{-2}$, which includes a close-up of Fig. 1(B) at 10 K. A pronounced suppression of the resistance is shown around there. Fig. 3(B) shows the magnetoresistance, $R_{xx}(B)$, at 10 K with $n = -3.48 \times 10^{12}$ cm$^{-2}$, which shows qualitatively the same behavior in the SC dome at lower temperatures. The data indicate that the SC dome exhibits rigidity under a magnetic field applied perpendicular to the substrate.

In conclusion, novel atomic-layer SC is discovered in graphene superlattices. SC via doping Dirac materials, in particular graphene superlattices, can be ubiquitous in condensed matter. Full details including more examples with high-$T_c$ are left to be studied beyond carbon-based materials.

**Materials and Methods**

The device fabrication via a modified dry-transfer technique is detailed in ref. [*12*, *13*] for hBN/BLG/hBN superlattices. Fig. 1(A) shows the schematics of our devices, wherein BLG is encapsulated between two hBN layers. The thickness of both the top and bottom layers of hBN is 30 nm. We fabricated the device by transferring BLG and hBN flakes onto an hBN substrate supported on a SiO$_2$/Si wafer. The sample was then etched into the H-bar geometry. The one-dimensional Cr/Au (= 5/55 nm, 0.5 μm × 0.5 μm) contacts were then deposited by electron beam (EB) evaporation followed by EB lithography. Note that the contact itself is non-superconducting without proximity effects. The sample quality of this device was estimated in ref. [*13*]. Herein, we show an example (see also the Supplementary Information). The quality of the graphene-related device is closely related to δ$n$, indicating the sharpness of the resistance peak at CNP/DP. In our device, δ$n$ is less than $5 \times 10^{10}$ cm$^{-2}$, which is comparable to that in ref.

[*21*]. Since the report of ref. [6], the role of disorder and inhomogeneity has been reported in graphene SC, and high-quality device should be crucial to realize SC (see the Supplementary Information for the sample quality). For hBN/SLG/hBN superlattices in the Supplementary Information, see ref. [*12*] for details and Fig. S2(A) for schematics. The thickness of the top and bottom hBN layers is 16 nm and 20 nm, respectively. Note that Hall-bar geometry is employed in this case.

Measurement setup is as follows. For hBN/BLG/hBN superlattices, the resistance is defined through the four-terminal resistance, $R_{ij,kl}$, which is defined by the voltage drop between terminals $k$ and $l$ divided by the electrical current injected between $i$ and $j$ (see also Fig. 1(A)). We have also checked the two-terminal measurement and the shuffling of terminals, which provides consistent results. All electrical contacts are ohmic down to the lowest temperatures. The measurement was performed at 1.5 K–80 K using both DC and a low-frequency (17 Hz) lock-in technique with an AC excitation current of 10–100 nA and in variable temperature cryostats (two types of cryostats having base temperatures of 5 K and 1.5 K were used). $R_{12,43}$ defines the longitudinal resistance $R_{xx}$. $R_{xy}$ is defined by $R_{13,42}$. Our device shows four-terminal rectangular structures, wherein the mean free path is within the device dimensions (the order of 1 μm) [*13*] and the I–V characteristics reveal an ohmic behavior in the normal phase [*22*]. $R_{13,42}$ leads to the Hall coefficient after a symmetrization as a function of magnetic field $B$ [*23,24*]. The Hall conductance $\sigma_{xy}$ is defined by $R_{xy}/((R_{xx}W/L)^2 + R_{xy}^2)$. For the measurement setup the hBN/SLG/hBN superlattice in the Supplementary Information, see ref. [*12*] for details.

**Acknowledgments:**

We thank Kensuke Kobayashi of Osaka University and Akinobu Kanda of Tsukuba University for helpful discussions. The device fabrication and measurement were supported by the Japan Society for Promotion of Science (JSPS) KAKENHI 26630139; and the NIMS Nanofabrication Platform Project, the World Premier International Research Center Initiative on Materials Nanoarchitectonics, was sponsored by the Ministry of Education, Culture, Sports, Science and Technology (MEXT), Japan. S.M. acknowledges financial support from a Murata Science Foundation. Growth of hexagonal boron nitride crystal was supported by the Elemental Strategy Initiative conducted by the MEXT, Japan and the CREST (JPMJCR15F3), JST.

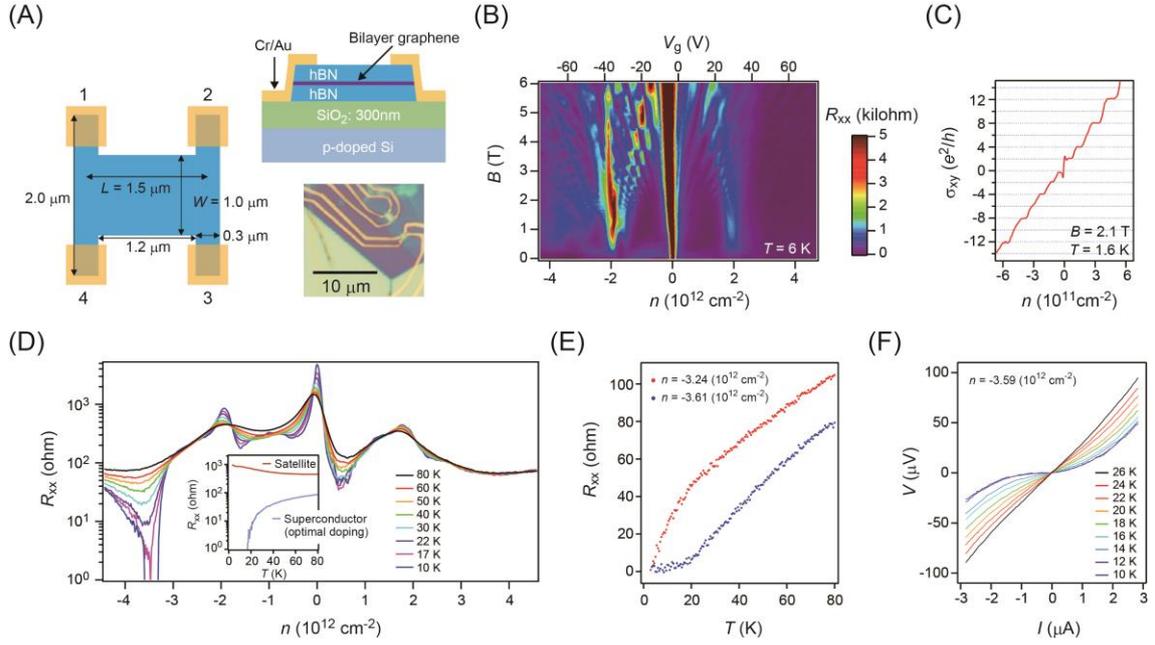

**Fig. 1. Characterization of our devices.** (**A**) Schematic of our hBN/BLG/hBN superlattice and the four-terminal measurement scheme. The one-dimensional Cr/Au (= 5/55 nm, 0.5 μm × 0.5 μm) contacts were deposited. *In situ* tuning of the electron density was performed via a back gate beneath the bottom hBN layer. An optical microscope image of the two devices is also shown. The larger of the two is employed in this paper, which corresponds to the schematic. (**B**) Intensity map of the longitudinal resistance, $R_{xx}$, as a function of the gate voltage $V_g$ and the magnetic field, $B$ (applied perpendicular to the substrate), at 6 K. The correspondence is also shown between the gate voltage, $V_g$, and density, $n$. (**C**) Quantum Hall effect occurs in our device at $T$ = 1.6 K and $B$ = 2.1 T. The Hall conductance ($\sigma_{xy}$) in steps of $4e^2/h$ are shown versus the density, $n$, which are the characteristics of BLG. (**D**) The resistances, $R_{xx}$, as a function of $n$ at $B$ = 0 T at various temperatures. The inset provides $R_{xx}$ as a function of $T$ at the satellite resistance peak ($n = -1.93 \times 10^{12}$ cm$^{-2}$) and the optimal doping for SC ($n = -3.48 \times 10^{12}$ cm$^{-2}$). (**E**) The resistances, $R_{xx}$, as a function of temperature, $T$, at different densities, $n$, with $B$ = 0 T ($n = -3.24, -3.61 \times 10^{12}$ cm$^{-2}$). (**F**) The *I–V* characteristics are shown at various temperatures with $B$ = 0 T near the optimal doping ($n = -3.59 \times 10^{12}$ cm$^{-2}$).

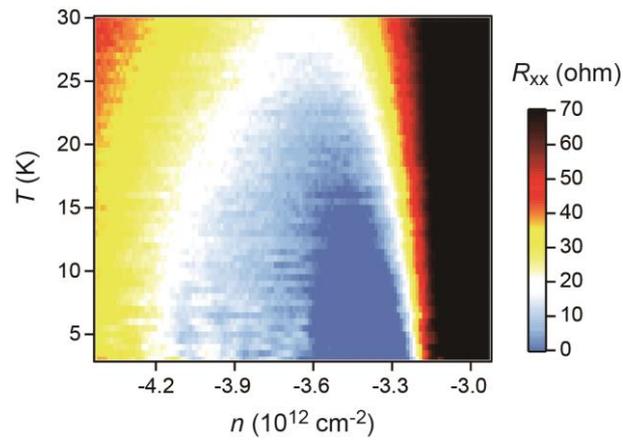

**Fig. 2. *In situ* electrostatic doping and the SC dome in our devices ($B = 0$ T).** The resistance, $R_{xx}$, as a function of both the density, $n$, and temperature, $T$. An SC dome is shown (blue-colored region).

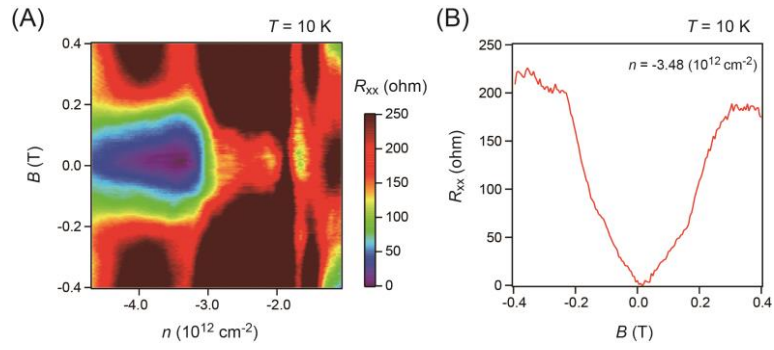

**Fig. 3. Magnetoresistance in our devices.** **(A)** An intensity map of the longitudinal resistance, $R_{xx}$, as a function of the density, $n$, and the magnetic field, $B$ (applied perpendicular to the substrate). This includes a close-up of Fig. 1(B) with a focus on the regime near $n \sim -3.5 \times 10^{12}$ cm$^{-2}$ at $T = 10$ K. **(B)** The magnetoresistance, $R_{xx}(B)$, at $T = 10$ K with $n = -3.48 \times 10^{12}$ cm$^{-2}$.

# Supplementary Information for

## Observation of superconductivity in bilayer graphene/hexagonal boron nitride superlattices

Satoshi Moriyama[*], Yoshifumi Morita[*], Katsuyoshi Komatsu, Kosuke Endo, Takuya Iwasaki, Shu Nakaharai, Yutaka Noguchi, Yutaka Wakayama, Eiichiro Watanabe, Daiju Tsuya, Kenji Watanabe, Takashi Taniguchi

Correspondence to: MORIYAMA.Satoshi@nims.go.jp (S.M.); morita@gunma-u.ac.jp (Y.M.).

**Supplementary Text**

**#1 Sample quality evidenced by the estimation of charge inhomogeneity**

The quality of our device, as the upper bound of the charge inhomogeneity at the CNP is estimated to be $\delta n$, is less than $5 \times 10^{10}$ cm$^{-2}$ from the peak width of the CNP, as in Ref [21] (Fig.S1).

**#2 A comparative study: Single-layer graphene superlattices**

In the main text, our focus is on *in situ* electrostatic doping in bilayer graphene (BLG). For a comparative study, here we study single-layer graphene (SLG) superlattices.

The emergence of Dirac fermions in solids ("Dirac materials") has been well established since the discovery of "1st generation" in graphene [4, 5]. Here, we focus more on emergent Dirac fermions in solids, i.e., we search for "higher generations".

Our focus is on hexagonal boron nitride (hBN)/SLG/hBN structures, which harbor higher-generation Dirac fermion points with a narrow bandwidth, i.e., with a relatively strong correlation.

Our device was fabricated by stacking SLG and two thin sheets of hBN in an hBN/SLG/hBN stack with a small angle between one of the two sheets of hBN and the graphene (Fig. S2(A)). This superlattice provides a stage for our demonstration. In monolayer graphene with Dirac-type relativistic energy dispersion, the inversion symmetry can be broken by stacking graphene on an hBN substrate with an angle near zero degrees, which leads to a long-length Moiré pattern due to the 1.8% lattice mismatch between the graphene and the hBN. We fabricated hBN/SLG/hBN heterostructures with one-dimensional Cr/Au contacts. Fig. S2(A) shows a schematic of the typical structure of our device. $R_{63, 54}$ defines the longitudinal resistance $R_{xx}$. $R_{63, 51}$ defines a Hall resistance. Fig. S2(B) shows an intensity map of the longitudinal resistance $R_{xx}$ as a function of the back-gate voltage, $V_g$, and the magnetic field, $B$ (applied perpendicular to the substrate), at 1.5 K. Sharp increases in the longitudinal resistance ($R_{xx}$) at $V_g$ values of 0 V and −21 V corresponding to the 1st (DP) and 2nd (SDP) generation Dirac points, respectively. The emergence of the SDP is a consequence of energy band engineering due to the misalignment of the graphene and hBN crystals, which leads to energy gaps at DP and SDP. The Quantum Hall effect (QHE) of SLG is observed near the DP, and a Landau-fan diagram is observed [*17*].

Upon *in situ* electrostatic doping away from the SDP, we observed "signatures" of SC. Fig. S3(A) is a zoom-in of Fig. S2(B) with a focus on the regime near $V_g \sim -26.5$ V. Pronounced suppression of the resistance is shown around there, which resides near the van Hove singularity (vHs), and the sign of the carrier charge changes (but it is not DP/SDP), as discussed below. Fig. S3(B) shows the magnetoresistance at $T = 1.5$ K with $V_g = -26.55$ V near the vHs. We confirmed that this device did not show an SC dome in the dilution refrigerator measurement of the mixing chamber temperature $T_{mix} = 40$ mK and that it is weakly SC at best.

**#3 Estimation of the carrier density via the low-field Hall effect**

In Fig. S4, the carrier density, $n_H$, is shown as a function of the gate voltage, $V_g$. With the carrier density, $n_H$ is estimated via the low-field Hall effect. In the low-temperature limit, we estimated $n$ using extrapolating a linear relation ($n_H$ versus $V_g$) from CNP/DP, which was consistent with the estimation through our device structure/electrostatic capacitances.

Note that, in the beginning, the sign of the carrier changes at higher-generation Landau fans due to switching from electrons to holes. In the following, we focus on the sign change of the carrier away from such points.

In the case of SLG (Fig. S4(A)), the longitudinal resistivity, $\rho_{xx}$, shows a dip structure (some "signature") in the vicinity of the point where the sign of the carrier changes (however, it does not belong to the 1st or 2nd generations of the Landau fans), i.e., near vHs. The dip is located at $V_g = -26.55$ V with $\rho_{xx} = 1.6$ ohm.

In the case of BLG (Fig. S4(B)), as $V_g$ is swept, the sign of the carrier changes as implied by the low-field Hall effect. The SC dome resides near such sign-changing points, as discussed in the main text.

#4 Temperature dependence of the resistivity: Global picture

In the main text, our focus is on the SC dome and the low-temperature regime. Herein, as a compliment to the main text, we discuss the temperature dependence of the resistivity from a more global point of view.

In Fig. S5, the longitudinal resistances are shown as a function of the temperature, $T$, for the case of both SLG (Fig. S5(A, B)) and BLG (Fig. S5(C, D)). Herein, we comment on several scenarios for the exponent, $\alpha$, i.e., $\Delta R(T) \sim T^\alpha$, which is a temperature-dependent part of the resistance with the residual resistance subtracted. $\alpha = 2$ is a result of the celebrated Fermi-liquid exponent, which was recently assigned to the Umklapp process in graphene superlattices [25]. $\alpha = 1$ is reminiscent of "strange metal" in cuprates [26]; however, it is also consistent with scattering by acoustic phonons [27, 28]. Furthermore, we assume a two-fluid model due to nodal and antinodal components. This is reasonable in some graphene superlattices (and some materials with unconventional density waves). This two-fluid model leads to a crossover between $\alpha = 1$ and 2 due to the Umklapp process [29]. A more careful assignment of the scattering mechanism will be discussed in a separate paper.

More comments are in order of temperature dependence. Triggered by the recent discovery of magic-angle SC [6], new players (SC and magnetism) have entered the stage for graphene superlattices. We believe that this is more ubiquitous than expected. Actually, we have encountered many "signatures" in more general settings without "magic". Herein, we show an example. In Fig. S5(C, D), the temperature dependences of the longitudinal resistance between CNP and the satellite is also shown for BLG ($V_g$ = −17.4, −28.0 V). This indicates non-metallic behavior between CNP and the satellite. In our device, however, we confirmed that this does not lead to the "Mott" insulator or SC even at $T_{mix} = 40$ mK.

#5 Transition temperatures analysis

Some comments are for fixing transition temperatures (Fig.S6). For the analysis of the critical temperature, we applied Berezinskii–Kosterlitz–Thouless (BKT) analysis, where $ln(R_{xx}(T)/R_0) = -b(T/T_{BKT} - 1)^{-1/2}$ was assumed ($R_0$ and $b$ were non-universal, material-dependent parameters) [30,31,32]. The result is consistent with $T_{BKT}$, which is deduced from the $I$–$V$ characteristics (see also below). Now let us examine transition temperatures of atomic-layer SC's. In the beginning, focus on two characteristic regimes in 2D SC, one is the fluctuation regime near $T^*$, at which the amplitude of the order parameter develops; however, the superfluid density is renormalized to zero. The other is the SC regime below the BKT transition temperature, $T_{BKT}$. Here let us define the temperature $T^{onset}$ at which SC onsets, i.e., 90% of the total transition. In our case, the normal resistance shows a linear behavior with $T$ [26,33]. We define the normal resistance for the definition of $T^{onset}$ by the value where it deviates from the linear form. We define $T^{**}$ as 50% of the total transition. In our SC device, for example in Fig. S6(A, B)), $T^{onset} \sim 50$ K, $T^{**} \sim 30$ K and $T_{BKT} = 14$ K. As discussed below in this section, via an analysis of the excess conductivity due to SC fluctuations, the crossover temperature, $T^*$, at which the finite amplitude of SC order develops is estimated, which is close to $T^{**}$ in many SC's, and we sometimes identify the two ($T^*$ and $T^{**}$) as a crossover temperature.

Complementary to above, we analyze the characteristic temperature by excess conductivity due to SC fluctuations. For the resistance, $R_{xx}(T) = (1/R_N(T) + (W/L)\Delta G)^{-1}$ is assumed. The normal resistance $R_N(T)$ is set to be in a $T$-linear form $a + bT$, which has been observed ubiquitously near the SC domes of graphene superlattices [26,33]. The excess conductivity due to SC fluctuations, $\Delta G$, comprises two terms the Aslamazov–Larkin (AL) term [14] and the Maki–Thompson (MT) term [15,16], where we introduce the pair-breaking parameter $\delta$. Including the depression of the electronic density of states due to SC fluctuations, $\Delta G = \Delta G_{AL} + \Delta G_{MT} + \Delta G_{DoS} = (e^2/16\hbar)T^*/(T - T^*) + (e^2/8\hbar)[T^*/(T(1 - \delta) - T^*)] \ln[(T - T^*)/\delta T]$ [34]. An example is shown below in Fig.S6 (B, C).

Further, we deduce $T_{BKT}$ from the I–V characteristics. Near the BKT transition, $V \sim I^\alpha$ with $\alpha = 3$ at $T = T_{BKT}$ [35,36]. Finite size corrections smear out the "universal jump" from $\alpha = 1(T > T_{BKT})$ to $3(T = T_{BKT})$. Above $T_{BKT}$ ($T > T_{BKT}$), deconfined vortices lead to ohmic resistance ($\alpha = 1$). Below $T_{BKT}$ (for $T \leq T_{BKT}$), confinement of vortex–antivortex pairs occurs. Applying a finite bias current, current-induced free vortices dominate here, which leads to the scaling law for the non-linear current–voltage curve discussed above. Furthermore, nonreciprocity due to hBN-induced inversion-symmetry breaking can play some role there in our device. An example is shown below in Fig. S6(D, E, F).

## #6 Another device

We also studied another H-bar device to check/exclude the role of artificial geometrical effects. In Fig. 1(A), the optical microscope image of two devices is shown. In the main text, we employed the larger of the two. We show typical data for the smaller one, where the onset of SC is reconfirmed with an approximate particle–hole symmetry of CNP (Fig.S7).

#7 Magnetic field responses of the SC state

Stability under a magnetic field is a hallmark of superconductors (SCs), which implies "rigidity" of the condensate [37]. In the SC regime, where the rigidity develops, a vortex state can emerge under a magnetic field. Herein, the vortex motion is a relevant origin for the voltage drop due to the Josephson relation. Note that vortex motion in quantum-limited ($k_F\xi$~1, $k_F$ Fermi momentum, $\xi$ superconducting coherence length) SC combined with the extremely clean (xC) limit has not been investigated in detail where extrinsic effects, e.g., disorder and/or random pinning, are suppressed. Further, in SC with broken inversion symmetry, e.g., due to hBN, vortices moving via an external electric current, can feel an effectively asymmetric potential, i.e., a "ratchet" effect due to nonreciprocal SC. Although the vortex phase diagram and its details remain to be examined for our device, we show preliminary phase diagrams (Fig. S8). The inset of Fig. S8 shows resistance as a function of the magnetic field, $B$, applied perpendicular to the substrate at several temperatures. Although the boundary between the normal state and the vortex/mixed state can be a crossover, we define $B^*$ by the magnetic field at which 50% of the "normal" resistance is recovered just for convenience. Here some comments are in order on the "normal" resistance. We define it by the regime where the precursor to SC, i.e., the large magnetoresistance is sufficiently suppressed, as temperature is raised. Note that, in our SC dome, the metallic regime can be smoothly connected to the insulating regime as the magnetic field is increased at finite temperature (the inset of Fig. S8).

The $B^*$ is fitted to the orthodox theory (Fig.S8) [38], but detailed study on the vortex phase diagram is left for future work, e.g., crossover to the QHE and "creepy" effects. In particular, our device can be in the vicinity of quantum-limited SC ($k_F\xi$~1) combined with the xC limit, which also implies a large Maki parameter $\alpha_M$ [39] and large Pauli-paramagnetic effects.

#8 Quantum phase-coherent transport of the SC state

Evidence of a percolating superconducting/Josephson junction network has recently been suggested in the SC by Ref. [6]. Figure S9 shows periodic oscillations in the $I$–$V$ characteristics of our SC devices, implying Fraunhofer interferences. As proposed in Ref. [6], we believe that this is a result of quantum phase-coherent transport in the Josephson junction (JJ), which is self-organized in our SC devices due to the extrinsic disorder and/or intrinsic characteristics of mesoscopic SC (with competing orders). Even though the details of the JJ are unknown and a quantitative description is not straightforward, the period of the oscillations $\Delta B$ is ~10 mT, which leads to an effective JJ area of $S$ ~ 0.1 μm$^2$. Here, $S = \Phi_0/\Delta B$ ($\Phi_0 = h/2e$, where $h$ is Planck's constant and $e$ is the electron charge). This JJ can provide a phase-sensitive test of the pairing symmetry [40]. The Fraunhofer pattern in our devices implies an analog of the so-called π junction and unconventional superconductivity [41,42]. Another scenario for an analog of the π junction is the presence of magnetic order. We note that magnetism due to correlated insulator behavior is not naively expected in the focused regime of our device in the main text. However, a more careful analysis is left as future work, e.g., via magnetic spectroscopy. For example, even though there is no signature of correlated insulator behavior in the device in the main text, our device can be in the vicinity of criticality of, for example, (unconventional) charge/spin/valley density waves (c/s/vDW) or a "Mott" insulator [6], as implied by another H-bar device in the Supplementary Information.

This Fraunhofer pattern has its origin in the inhomogeneity of our device in the low-temperature regime [6] and should be distinguished from other commensurability patterns due to, for example, ballistic channels. We confirmed such a pattern away from the SC dome, e.g., magnetic focusing [43], which is absent near the SC dome.

Finally, note that inhomogeneous states should occur at the low-temperature limit as indicated in this section. Providing a consistent picture of this regime including electrical contacts is an open problem and is beyond a simple BKT analysis.

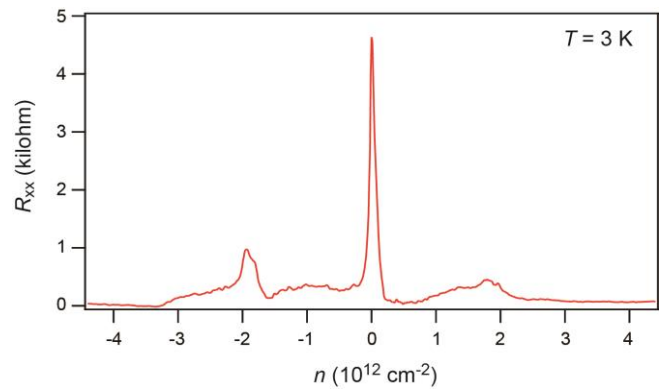

**Fig. S1. Resistance peak at the CNP of our BLG superlattices.** $R_{xx}$, as a function of the density, $n$, at 3 K. The quality of our device (residual carrier density) is estimated based on the peak width of the CNP.

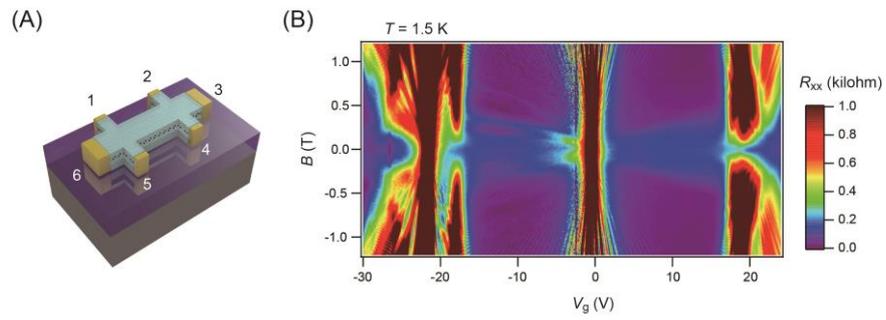

**Fig. S2. Characterization of our SLG superlattices.** **(A)** Schematic of our hBN/SLG/hBN superlattice with one-dimensional Cr/Au contacts and the four-terminal measurement scheme. **(B)** Intensity map of the longitudinal resistance, $R_{xx}$, as a function of the gate voltage, $V_g$, and the magnetic field, $B$, (applied perpendicular to the substrate), at 1.5 K.

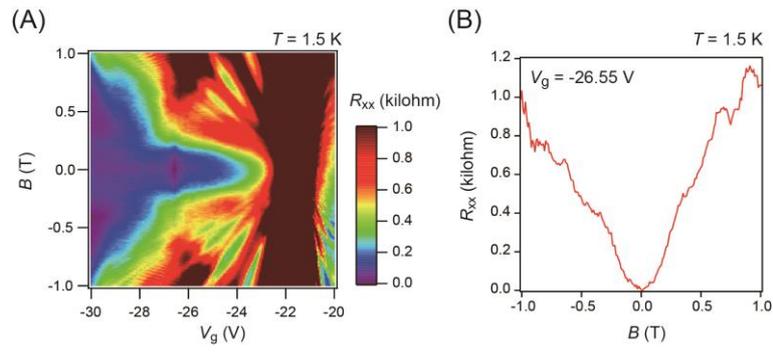

**Fig. S3. Magnetoresistance in our SLG superlattices.** (**A**) Close-up of Fig. S2(B) with a focus on the regime near $V_g \sim -26.5$V, which resides near a van Hove singularity (vHs). (**B**) The magnetoresistance, $R_{xx}(B)$, at 1.5 K with $V_g = -26.55$ V near the vHs.

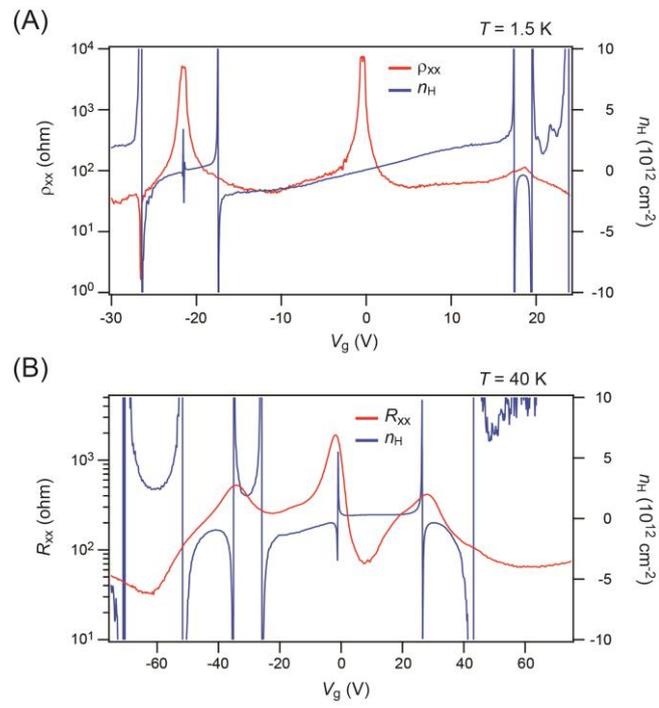

**Fig. S4. Carrier density estimated via the low-field Hall effect for our SLG/BLG superlattices.** The carrier density, $n_H$, is shown as a function of the gate voltage, $V_g$, for **(A)** SLG at 1.5 K (with the resistivity $\rho_{xx}$) and **(B)** BLG at 40 K (with the resistance $R_{xx}$).

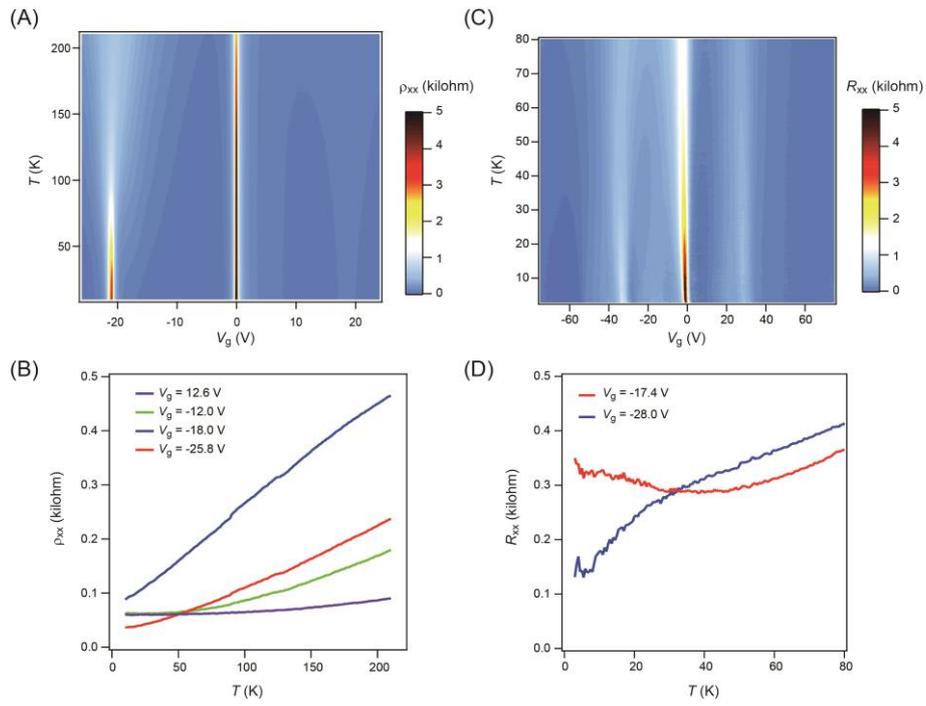

**Fig. S5. Temperature dependence of the resistance for our SLG/BLG superlattices**
The resistance as a function of the temperature, $T$, for some values of the gate voltage, $V_g$, for **(B)** SLG and **(D)** BLG. The mapping of the resistance is also shown as a function of $T$ and $V_g$ for **(A)** SLG and **(C)** BLG.

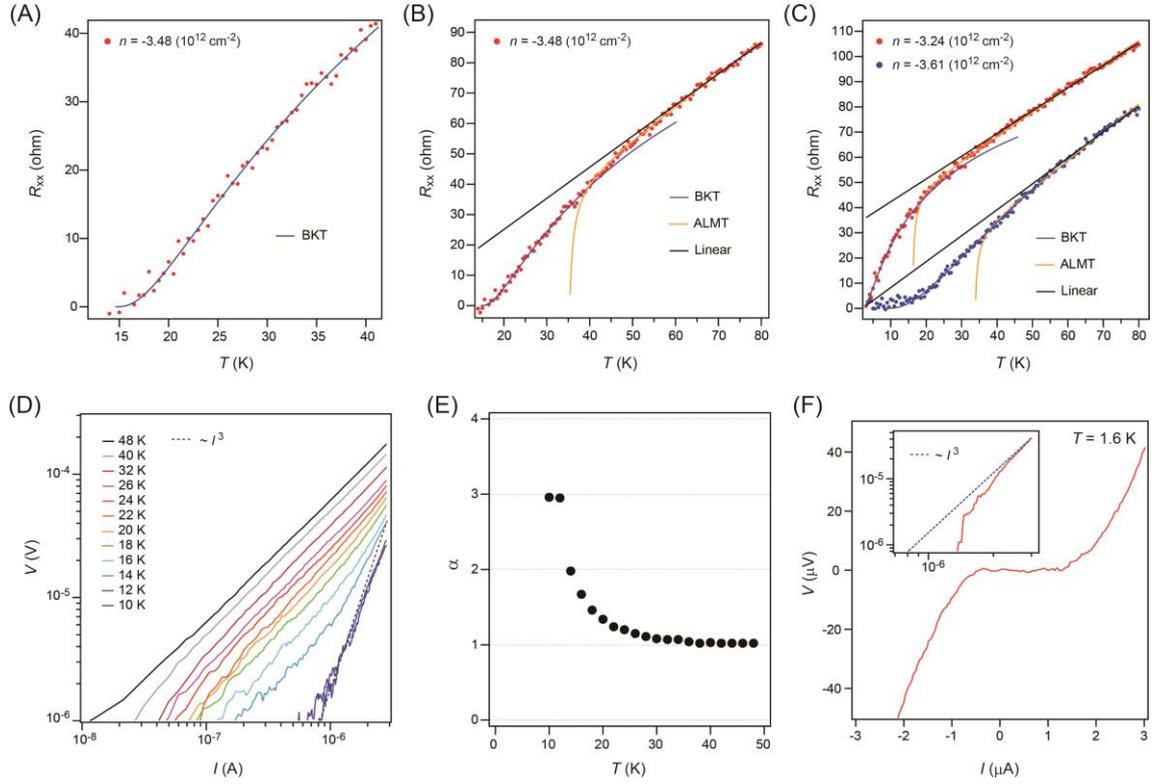

**Fig. S6. Transition temperatures analysis.** (**A**) An example ($n = -3.48 \times 10^{12}$ cm$^{-2}$ for our BLG superlattices) of the measured resistance, $R_{xx}(T)$, is shown (red points), which is consistent with the BKT-type analysis (blue line denoted by "BKT") and leads to an estimation of the $T_{BKT}$. In this case, $T_{BKT} = 14.3 \pm 0.9$ K, and it is recorded as $T_{BKT} = 14$ K. The broad transition character causes uncertainty in the fitting procedure and an error bar in fixing the $T_{BKT}$. (**B**) $T^{onset}$ and $T^*$ are also marked, where "ALMT" denotes a curve with SC fluctuations (described by the AL and MT terms). A fitting to the $T$-linear function ("Linear") is shown for the high-temperature regime, which acts as a guide for the eye. In this case, $T^{onset} \sim 50$ K and $T^* \sim 30$ K. (**C**) The same analysis as (B) is done for the data in Fig. 1(E). (**D**) $I$–$V$ characteristics are shown in log-scale for Fig. 1(F) for the BKT-type analysis. (**E**) Exponents ($V \sim I^\alpha$) are shown as a function of temperature $T$ for the data in (D). $\alpha = 3$ gives a consistent transition temperature $T = T_{BKT}$. (**F**) An example of $I$–$V$ characteristics at a low temperature $T = 1.6$ K. Inset: the same $I$–$V$ in log scale.

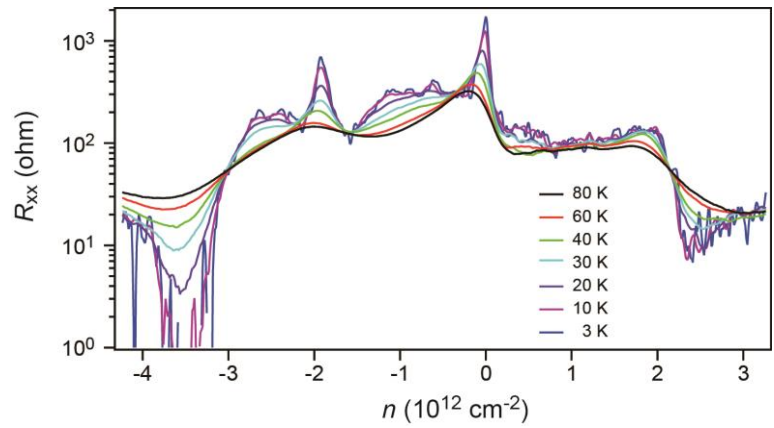

**Fig. S7. Another H-bar device.** We also conducted a study on another H-bar device to check/exclude the role of artificial geometrical effects. Typical data are shown, where the onset of SC is reconfirmed. The resistances, $R_{xx}$, are shown as a function of $n$ at $B = 0$ T at various temperatures.

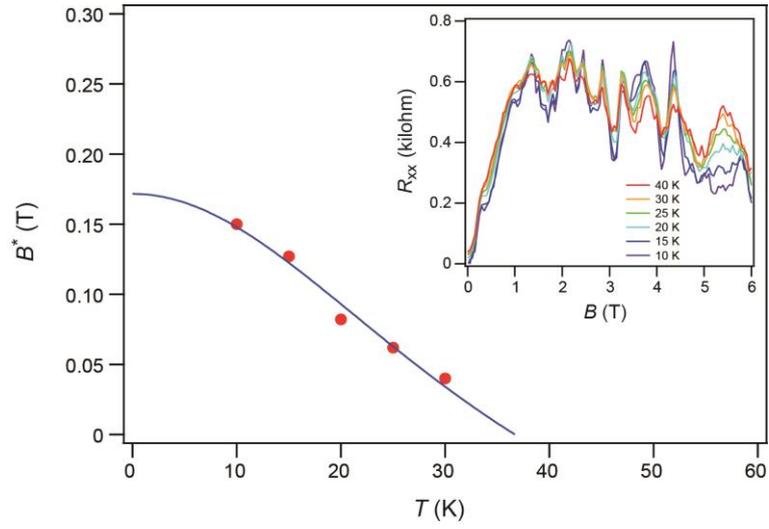

**Fig. S8. Magnetic field response of our devices.** An example of $B^*$ as a function of the temperature $T$ near the optimal doping ($n = -3.3 \times 10^{12}$ cm$^{-2}$) for our BLG superlattices with a different thermal cycle from the main text. The curve is the Ginzburg-Landau form $\sim(1- (T/T^*)^2)/ (1+ (T/T^*)^2)$ [38]. In this case, $T^* = 36.7 \pm 1.5$ K. Inset: Resistance as a function of the magnetic field. $B$, applied perpendicular to the substrate at various temperatures.

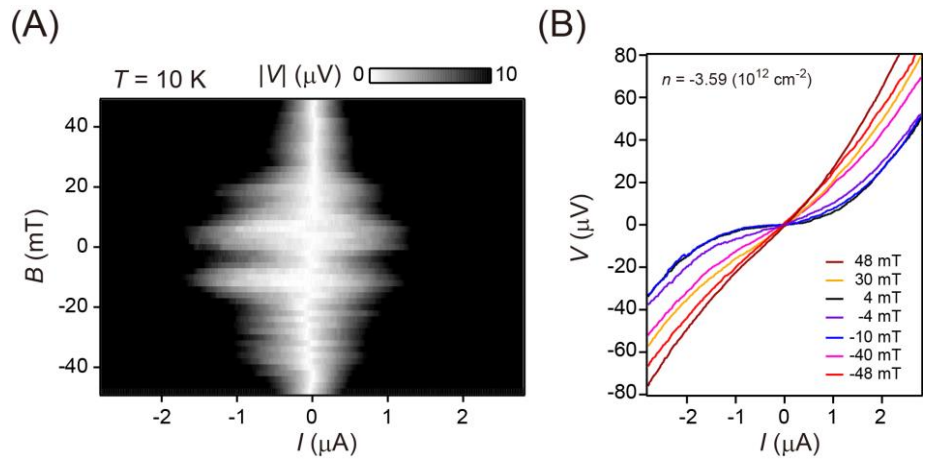

**Fig. S9. An example of quantum phase-coherent transport in our devices.** **(A)** A gray scale plot of the measured $V$ as a function of $I$ and $B$ at $n = -3.59 \times 10^{12}$ cm$^{-2}$ and $T = 10$ K with a different thermal cycle from that in the main text. Periodic oscillations in the $I$–$V$ characteristics are observed, implying Fraunhofer interference. The magnetic field $B$ is applied perpendicular to the substrate. **(B)** Typical $I$–$V$ characteristics under the magnetic field in panel (A).